\documentclass{bmcart}

\usepackage[latin1]{inputenc} 
\usepackage{graphicx}
\usepackage{datetime}
\usepackage{caption}
\usepackage{subfigure}
\usepackage{float}
\begin{document}

\begin{frontmatter}

\begin{fmbox}

\title{A Comparison of Spatial-based Targeted Disease Containment Strategies using Mobile Phone Data}


\author[
   addressref={aff1},                   
   corref={aff1},                       
   email={stefania.rubrichi@orange.com}   
]{\inits{SR}\fnm{Stefania} \snm{Rubrichi}}
\author[
   addressref={aff1},
   email={zbigniew.smoreda@orange.com}
]{\inits{ZS}\fnm{Zbigniew} \snm{Smoreda}}
\author[
   addressref={aff2},
   email={m.musolesi@ucl.ac.uk}
]{\inits{MM}\fnm{Mirco} \snm{Musolesi}}


\address[id=aff1]{
  \orgname{SENSE,Orange Labs}, 
  \street{44 avenue de la République},                     %
  \postcode{92326}                                
  \city{Chatillon},                              
  \cny{FR}                                    
}
\address[id=aff2]{%
  \orgname{Department of Geography, University College London},
  \street{Grower Street},
  \postcode{WC1E 6BT}
  \city{London},
  \cny{UK}
}

\end{fmbox}


\begin{abstractbox}

\begin{abstract} 
Epidemic outbreaks are an important healthcare challenge, especially in developing countries where they represent one of the major causes of mortality. Approaches that can rapidly target subpopulations for surveillance and control are critical for enhancing containment and mitigation processes during epidemics.

Using a real-world dataset from Ivory Coast, this work presents an attempt to unveil the socio-geographical heterogeneity of disease transmission dynamics. By employing a spatially explicit meta-population epidemic model derived from mobile phone Call Detail Records (CDRs), we investigate how the differences in mobility patterns may affect the course of a hypothetical infectious disease outbreak. We consider different existing measures of the spatial dimension of human mobility and interactions, and we analyse their relevance in identifying the highest risk sub-population of individuals, as the best candidates for isolation countermeasures.
The approaches presented in this paper provide further evidence that mobile phone data can be effectively exploited to facilitate our understanding of individuals' spatial behaviour and its relationship with the risk of infectious diseases' contagion. In particular, we show that CDRs-based indicators of individuals' spatial activities and interactions hold promise for gaining insight of contagion heterogeneity and thus for developing mitigation strategies to support decision-making during country-level epidemics.

\end{abstract}


\begin{keyword}
\kwd{spatial networks}
\kwd{mobile phone data}
\kwd{human mobility}
\kwd{epidemic spread}
\end{keyword}


\end{abstractbox}
%

\end{frontmatter}



\section*{Introduction}
Epidemic outbreaks represent an important healthcare challenge, especially in developing countries where they represent one of the major causes of disease suffering and mortality. For this reason, an in-depth understanding of epidemic transmission dynamics on a countrywide scale is critical in elucidating, facing and controlling epidemics. Disease spreading is a highly heterogeneous process, with certain areas (or indeed individuals) being at higher-risk than others. Therefore, drastic population-wide measures, like quarantining entire countries, are often ineffective, at times harmful~\cite{meloni,chamary}, as well as costly and difficult to implement. Recently, it has been shown that improvements may be achieved through targeted control strategies~\cite{lloydsmith,lima1,halloran}. Individual variation in rates of infectious contact can significantly alter patterns of disease spread~\cite{merler,lloydsmith}. This calls for an in-depth and systematic investigation of such heterogeneity.

In this work we consider person-to-person, directly spread infectious disease epidemics, where transmission occur because of individuals' co-location and/or face-to-face interactions. We simulate the dynamics of a disease outbreak and explore the effects of targeted mitigation strategies. For these diseases spatial propagation is largely dependent on human mobility. People move across several locations, both exposing themselves to infectious agents in these locations and transport these agents between them. Therefore, real-world and fine-grained data on human mobility patterns and interactions are key elements for building effective epidemiological models~\cite{colizza1}. Furthermore, they may serve as informative surrogate to correlate infectiousness heterogeneity: systematic variations in mobility patterns of the population are sufficient to drive non-negligible differences in infectious disease dynamics~\cite{dalziel}.
Yet, access to highly detailed and updated data on population movement may be difficult and costly, especially when dealing with daily movements in small countries or at regional scale. Up to the last five years, the main sources of travel information have come from direct observations, census data and surveys~\cite{brown,lynch,stoddard}, which are sometimes scarcely applicable to large-scale studies, since they are too specific to be replicated generally~\cite{oreilly}.

More recently, mobile phone data have been made by cellular operators and, in particular, call detail records (CDRs). These are data collected from telecommunication companies for billing purposes, coming thus without extra cost or overhead, providing detailed temporal and spatial information about millions of cellphone users at various scales.
CDRs can be used to gather fine-grained information about individuals both in terms of mobility and, indirectly, their social network through their phone calls. Recent studies have explored the use of CDRs to quantitatively understand human mobility dynamics~\cite{gonzalez,simini} and all social activities and phenomena driven by it~\cite{blondel,calabrese}, including urban planning~\cite{louail}, emergency response~\cite{gundogdu} and, most importantly for the aim of this paper, epidemics control \cite{tizzoni,lima1}. In this regard, an important line of research has explored the use of CDRs for building epidemiological models of disease spreading. Proposed models range from approaches that consider aggregated flows to finer-grained meta-population or agent-based models \cite{wesolowski,colizza,tizzoni,lima1,lemenach}. Given the known correlation between proximity and social links \cite{lambiotte,onnela}, these models have been used to evaluate the influence of travel behaviour on spreading of diseases, to identify hotspot areas and to study diseases' containment strategies. However, only a few of these approaches have explicitly considered the spatial structure of the population~\cite{meloni}. It is well established that the spatial structure of the population has an impact on the diffusion of epidemics~\cite{merler}. 

Starting from this body of work, in this paper, we propose to investigate the correlation between the spatial dimension of individuals' travel behaviour  and epidemic diffusion, focussing on the quantification of the risk of infectiousness/infection of the population. In particular,  we explore and compare the effects of different targeted mitigation strategies based on the analysis of mobile phone data. Starting from~\cite{lima1}, we adopt a spatially-explicit transmission model in the form of a meta-population model. Meta-population models are used to describe disease spreading among several sub-populations that are spatially structured, and connected by a mobility network whose links denote individuals' moving across sub-populations. In each subpopulation disease contagion is modelled using a SEIR (susceptible-exposed-infected-recovered) compartmental model~\cite{keeling}. For the construction of our mobility network we use an anonymised CDR dataset about mobile phone usage in Ivory Coast containing billing information of about 8 million users collected over a nine-month period.

Given the dynamics simulated by the model, we explore and compare the effects of different targeted mitigation strategies that rely on the characterisation of the spatial behaviour of individuals. More specifically, by considering strategies both at geographical as well as individual level, we investigate the chance of success when targeting either higher-risk geographical areas or higher-risk individuals based on spatial characteristics of the mobility network as well as behaviour  to identify the best candidates for isolation.  More in general, the goal of this paper is to show that quantifying the role of space in mobility analysis will improve our understanding of diffusion processes. We will also provide evidence that successfully performing epidemic mitigation strategies may require the identification of differences in mobility patterns among individuals.

\section*{Materials and methods}
\subsection*{Data}
The empirical evaluation of this work is based on mobile phone and epidemiological data. We analysed an anonymised set of mobile phone data collected by Orange C\^{o}te d'Ivoire. It consists of billing information of about 8 million mobile phone users (i.e., 35\% of the country population), collected between February and October 2014 in Ivory Coast, for a total of about 4.5 billion records. Mobile phone operators continuously collect such data for billing purposes and to improve the operation of their cellular networks. Every time a person uses a phone, makes a call, sends an SMS or goes online, a Call Data Record is generated. The record contains the caller and callee IDs, timestamp, duration and type of communication, as well as an identifier of the cellular tower that handled the call. The approximate spatio-temporal trajectory of a mobile phone and its user can be reconstructed by linking the CDRs associated with that phone with the geographic location of the cellular towers that handled the calls.

As far as the epidemiological data is concerned, in order to place our results in a more realistic context, we consider a scenario modelled using values of the parameters estimated from the Ebola outbreak in Sierra Leone in 2014~\cite{althaus} (Tab. \ref{table1}). This type of modeling can be used for analyzing different ``what-if'' scenarios and for devising mitigation strategies. It is worth noting that we present the results considering a worst-case scenario, projecting the most severe form of Ebola epidemics.
\subsection*{Disease Spread Spatial Model}
In order to describe the countrywide-scale infectious disease spread, where individuals change location over time, we use a meta-population model. This framework has traditionally provided an attractive approach to epidemics modelling. In fact, a meta-population model allows modellers to include a realistic contact structure, and to reflect the spatial separation of the sub-populations (i.e., the contact rate might vary with spatial separation). The intuition behind meta-population models is that a natural population occupying any considerable area will be composed of a number $n$ of local populations (i.e., sub-populations), which interact and exchange individuals between them, because of their movement, through a given mobility network~\cite{andrewartha}. The nodes of such a network are the geographical areas connected according to a well-defined adjacency matrix $M$ (i.e., mobility matrix) of dimension $n$ by $n$. The element $m_{ij}$ represents the probability per unit of time that an individual chosen at random in an area $i$ will travel to an area $j$.

We compute this quantity using the CDRs dataset. Given users' movement trajectories, we estimate the probability of moving between antennas locations. A possible approach is to use a Markovian model as proposed in~\cite{lima1}. The estimation of the probability of movement is described by Equation~\ref{mobilityprobab}:

\begin{equation}\label{mobilityprobab}
m_{ij} = \frac{\sum_{u}M_{ij}^u}{\sum_{u}\sum_{k}M_{ik}^u}
\end{equation}

where $M_{ij}^u$ is the number of times an individual $u$ moves from an area $i$ to an area $j$. Daily location and movement are then aggregated to measure transitions among 508 Ivorian administrative regions called sub-prefectures.

Within each geographic area, sub-populations may be in contact and may change their health state according to the disease dynamics. By doing so, the system will evolve under the action of two processes, namely disease contagion and the mobility of individuals.

To model the process of disease transmission we consider the SEIR epidemiological model. Thus, in each node of the spatial network, SEIR dynamics takes place over a population of size $N_i(t)$ (the number of individuals located in an area $i$ at time $t$). With respect to the infection progress, individuals located in a given area $i$ are partitioned into $S_i(t)$, $E_i(t)$, $I_i(t)$, $R_i(t)$, denoting the number of susceptible, exposed, infected and recovered individuals at time $t$. Hence, at each time $t$, a person is either susceptible, exposed, infected or recovered (i.e., $S_i(t)+E_i(t)+I_i(t)+R_i(t) = N_i(t)$) and, as the SEIR process takes place, they change the state as follows: A susceptible individual becomes exposed to the disease with probability $\beta*I/N$, with $\beta$ being the product of the contact rate and the contagion probability. An individual that is exposed becomes infected at infection rate $\sigma$. An infected individual can then recover at a recovery rate $\gamma$. Finally or he/she can die before recovering because of infection-induced mortality with probability $\rho$~\cite{keeling}.

As stated above, simultaneously with the contagion process, individuals move according to the mobility matrix. So as time passes, $N_i(t)$ changes according to the number of individuals who have entered and who have left the node (i.e., geographical area) $i$, and the number of births and deaths. In order to combine the two interdependent processes and study their effect on the evolution of the system, we use the approach proposed by Lima et al.~\cite{lima1}, based on a product between the mobility matrix ($M$) transpose and the state variable vectors ($S$, $E$, $I$, $R$). Overall, the system can be described by the system of Equations \ref{seir}:

\begin{eqnarray}\label{seir}
S_i\bigl(t+1\bigr) &=& \sum_{j=1}^n m_{ji} \biggl[S_j(t) + \nu - \beta \frac{S_j(t)}{N_j(t)}I_j(t) -\mu S_j(t) \biggr]
\nonumber\\
E_i\bigl(t+1\bigr) &=& \sum_{j=1}^n m_{ji} \biggl[E_j(t) + \beta \frac{S_j(t)}{N_j(t)}I_j(t) -\sigma E_j(t) -\mu E_j(t)\biggr]
\nonumber\\
I_i\bigl(t+1\bigr) &=& \sum_{j=1}^n m_{ji} \biggl[I_j(t)+ \sigma E_j(t)  - \frac{\mu + \gamma}{1-\rho}I_j(t) \biggr]
\nonumber\\
R_i\bigl(t+1\bigr) &=& \sum_{j=1}^n m_{ji} \biggl[R_j(t) +\gamma I_j(t) -\mu R_j(t) \biggr]
\end{eqnarray}

where the expressions inside brackets describe the evolution of the disease according to the SEIR model, and the matrix product accounts for individuals moving between meta-populations. At each time step, individuals can change both state and location within the spatial network. Please note that this model takes into account also birth and mortality rates: these are modelled through the population level birth rate ($\nu$), and the per capita natural death rate ($\mu$).
\subsubsection*{Geographic-based Targeting}\label{geoIsol}
First, we consider spatial targeting. We approached this problem as the identification of influential spreaders within a complex spatial network. Traditional approaches to quantify the most efficient nodes in a network of interactions through which spreading processes take place have been based on centrality measures such as the degree, eigenvector centrality or k-shell~\cite{kitsak, borge, klemm}. These measures, although effective in identifying the most influential nodal position in a network, are rarely accurate in terms of the quantification of their spreading power of a given node, particularly for those that are not highly influential~\cite{lawyer}. 
This is because they are not able to capture and represent the dynamic processes that take place in the networked system under consideration (see for example the discussion in~\cite{borgatti}).

Fortunately, it has been showed that various approaches are effective in measuring node's influence in disease spreading processes. Here, in particular, we consider \textit{accessibility}, which has been shown to be effective in quantifying the relationship between structure and spreading dynamics~\cite{viana}. More specifically, this concept was introduced to quantify the efficiency of communications among nodes in a complex network. Several definitions of accessibility have been proposed. Our goal is to measure the possibility of interactions within an area. Thus, as suggested by Hansen~\cite{hansen}, we are interested in quantifying the inward accessibility, that is, for a given node $i$, the frequency of access to a node $i$ from all the other nodes of the network. 
For this reason, in order to quantify accessibility we adopted the \textit{place rank}~\cite{elgeneidy} measure. In particular, place rank is a flow-based accessibility measure, which uses origin-destination information to estimate the accessibility of a location within a geographic network. It is based on an intuition similar to that at the basis of Google Page Rank, i.e., the accessibility of a certain area is related to the probability of visiting it. For each node (area) of a network, it is determined considering the number of people moving to it. The contribution of the people of a certain area is a function of the accessibility of the area they come from and so on. More precisely, a place rank is defined following the algorithm presented below:

\begin{eqnarray}\label{place_rank}
P_{i,t} &=& \frac{R_{i,t}}{O_i}
\\
E_{ij,t} &=& E_{ij,t-1}*P_{i,t-1}
\\
R_{j,t} &=& \sum_{i=1}^I E_{ij,t}
\\
R_{i,t} &=& R_{j,t}^T
\\\nonumber
if R_{i,t} &=& R_{i,t-1}, stop; else: Eq. (3)
\end{eqnarray}

where $P_{i,t}$ is the power of the contribution of each person leaving $i$ at iteration $t$; $E_{ij,t}$ is the weighted origin-destination table, i.e. the weighted number of people leaving $i$ to reach $j$; $R_{j,t}$ is the place rank for zone $j$ at iteration $t$; $O_i$ is the number of people originating from $i$; $I$ is the total number of zones $i$ within the network.

\subsubsection*{Individual-based Targeting} \label{indivIsol}
We are aware that curbing the spread of a disease in an entire geographical region might be restrictive and somewhat difficult to implement. Thus, as a further improvement of the targeting process, we consider the ``spreading power'' of a single person based on their mobility profiles. We investigate the effect of specific spatial behavioural indexes, linked to users' mobility, on the identification of individuals at highest risk.
 
Studying human mobility and its relationships with people's daily activities might yield important insights into our understanding of human spatial behaviour.  In the past decade, human mobility has attracted large attention in several disciplines. One of the main findings is related to the spatial heterogeneity of human movement (see for example~\cite{gonzalez,song,pappalardo}).
We consider diversity of travel histories and mobility profiles, and try to link it to the heterogeneity of infectiousness levels. We propose to take into consideration the risk of infectiousness/infection of the population given individuals' travel behaviour.  The rationale is that the higher the mobility of an individual, the higher the probability to get infected, and if infected, to infect other individuals.

To this end, we analyse existing mobile phone-based mobility measures and study their correlation with the contagion risk of individuals. A significant body of literature has focussed on the characterisation of human mobility patterns as derived from CDRs data~\cite{gonzalez,song1,song,phithakkitnukoon}, resulting into the definition of several indicators for individual mobility. These indicators relate to certain extent to the different dimensions of mobility. In this work, we focus on measures that represent individual mobility from three critical perspectives: the spatial range (as measured by the radius of gyration), the spatial regularity (as measured by the movement entropy) and the percentage of time spent at home.

As an additional index for the quantification of contagion risk, we considered the hybrid $Progmosis$ risk model proposed by Lima et al.~\cite{lima}, which leverages both the mobility behaviour of single individuals and the epidemic dynamics itself.

We now discuss these indicators in more detail:
\paragraph{Radius of gyration:} it is one of the most frequently used measure for the characterisation of the spatial range of an individual $u$ and interpreted as the characteristic distance travelled by the individual  \cite{gonzalez,song1,lu,blumenstock,blumenstock1,wesolowski1,wesolowski2,wesolowski}. Given a spatio-temporal trajectory $M$, it measures the spatial spread of the visited locations in $M$ from the centre of mass of the trajectory (i.e., the arithmetic mean of the spatial locations in $M$). It is defined as:
\begin{equation}\label{radiusofgyration}
r_{g} = \sqrt{\frac{1}{N}\sum_{i \in L}n_i(r_i-r_{cm})^2}
\end{equation}
It is determined by first defining the geographic coordinates of the centre of mass $r_{cm}$ of all the $L$ locations $r_i$ visited by the individual. The straight-line distances from the centre of mass to each location are calculated, and the value of radius of gyration  is given by the square root of the mean of the squares of these distances. $n_i$  is the visitation frequency of location $i$, $N=\sum_{i \in L}n_i$ is the total number of visits.

\paragraph*{Movement entropy:} Besides the spatial range of mobility of an individual, we are also interested in considering its heterogeneity over the sequence of visited locations, by means of entropy. Entropy is a fundamental quantity, which is used to capture the degree of predictability of a time series~\cite{navet}. With respect to human mobility, it has been used to characterise its inherent predictability~\cite{song}. In particular, we adopted Shannon's entropy, defined as follows:
\begin{equation}\label{entropy}
S = -\sum_{i \in L}p_i \log p_i
\end{equation}
where $p_i$ is the historical probability that the location $i$ was visited by the user.

\paragraph*{Home staying:} It counts the percentage of interactions the user had while he was at home. We selected this spatial indicator as a measure of his/her homebound attitude capturing both the regularity (intended as the probability of finding the user in his most visited location) and the frequency of mobility. It is determined by first computing the position of user's home as the location where the user spends most of his time at night, then counting the number of calls the user makes from there.

\paragraph{Progmosis risk model:} Starting from the general definition of the risk associated to an event as the product of the event probability and the expected loss. Considering a disease with contagion rate per contact $\beta$ (i.e., given a friendship between an infected and a susceptible person, a contagion will happen with rate $\beta$); assuming the user $u$ spends $T_{u,l}$ fraction of his time in each location $l \in L_u$ (hence, $\sum_{i}T_{u,l}=1$), they define the contagion risk as:
\begin{equation}\label{progmosis}
C_u(t)= \beta \sum_{l,m \in L}T_{u,l} T_{u,m} (i_l(t) s_m(t) + i_m(t) s_l(t))
\end{equation}
where the probability of the event occurring is the probability that a person becomes infected in a region $l$, according to the time fraction spent there and the fraction of infected people $i_l$, while the expected loss is the number of people expected to be infected in another region, according to the time fraction spent there and to the fraction of susceptible people.

We used the bandicoot framework \cite{bandicoot} to extract the first three measures and we implemented the $Progmosis$ risk model. It is important to emphasise that with the term ``locations" here we refer to the Ivory Coast sub-prefectures.

\section*{Results}
Here we present the results using Montecarlo simulations of the model described above. We study the epidemic dynamics over time, considering three scenarios: (i) the total absence of mitigation measures for a period of seven months; (ii) the isolation of higher-risk areas; (iii) the isolation of higher-risk individuals. In each scenario, we extract patterns of individual mobility from CDRs on a daily basis, separated for weekdays and weekends, and obtain two matrices. Higher-risk areas as well as individuals to be isolated were selected according to the targeting strategies illustrated above, by using the CDRs data relative to the first five months of the dataset (form February 28, 2014 to August 15, 2014) to compute the spatial behavioural indexes described in \ref{geoIsol} and \ref{indivIsol} , and the remaining data (form August 16, 2014 to October 07, 2014) for the analysis of the evolution of the epidemics in presence of the mitigation strategies.
 
We first allocate the population of about $22$ million to the $508$ sub-prefectures over Ivory Coast according to the CDRs data. We then run $1000$ stochastic simulations, each one initialised with a small number of infected individuals in a randomly selected sub-prefecture used as a seed, corresponding to the 0.1\% of the entire population.
%
\subsection*{No Countermeasures Scenario}
We firstly explore the evolution of the epidemics in the case of absence of countermeasures. The average number of infected individuals over the whole seven-months observation period in this scenario is presented in Fig~\ref{figure2}.

%
\subsection*{Sub-prefecture-level Isolation Scenario (Geographic-based Targeting)}
In this scenario, we analyse the effects of quarantining a group of sub-prefectures selected using Place Rank and compare this strategy with a more traditional approach based on eigenvector centrality. To this end, we estimated the place rank values of each node (i.e., sub-prefecture) in the geographic mobility network. Then, in order to implement the quarantine strategies, we selected those with the highest values (i.e., top $1$, top $5$ and top $10$ highest ranked sub-prefectures) and curbed them by setting to $0$ the $i-th$ row and column of the mobility matrix, except for the elements $m_{ii}=1$.

Moreover, we also investigate the impact of timing of interventions over outcomes. Delay at which mitigation interventions are implemented is crucial for strategic epidemic control, but it may vary according to difficulties in identifying a novel outbreak, as well as other logistical, and economic constraints. To this end, we consider four scenarios for control planning: initiate the intervention (i) three, (ii) seven, (iii) ten, (iv) fourteen days after the infection starts.

Fig~\ref{figure3} shows that both centrality-based (left panel) and place rank-based (right panel) isolation strategies reduce the number of infections compared to the no countermeasure scenario. Moreover, the place rank-based metric outperforms the centrality-based one when isolating the top 5 and top 10 sub-prefectures as it is possible to observe in Fig~\ref{figure3.1}. As discussed above, the place rank indicator has been shown to be accurate in quantifying spreading power of nodes within a spatial network, especially for those that do not have an influential node position.

Concerning timing, as intuitively expected, results in Fig~\ref{figure3} indicate that the earlier an intervention is put in place the greater the beneficial effect in terms of total epidemics size. Thus, optimal mitigation options should be put in place as rapidly as possible.
%
\subsection*{Individual-level Isolation Scenario (Individual-based Targeting)}
Here we focus on the impact of individual behaviour on epidemics dynamics. We perform the simulations under six scenarios: (i) no countermeasures, i.e., the baseline scenario, (ii) isolating a portion of individuals randomly, (iii) isolating a portion of individuals with higher value of radius of gyration, (iv) isolating a portion of individuals with higher value of entropy of visited locations, (v) isolating a portion of individuals with lower value of home staying index, (vi) isolating a portion of individuals with higher value of $Progmosis$ risk model. The percentage of isolated individuals varies from 1\% to 10\% of the whole population with step length 1, and from 10\% to 30\% of the whole population with step length equal to 5. The intervention starts three days after infection.

From a practical point of view, individuals' isolation has been performed by removing their associated records from the whole dataset, and re-computing the probabilities $m_{ij}$.
Results are presented in Fig~\ref{figure4} in terms of total number of infected individuals over the time. The figure presents the results of simulations when isolating 1\%, 10\%, 15\% and 20\% of the whole population (see Additional Materials for more detailed results). Each scenario is represented by a colour; dotted lines are the associated 95\% confidence interval.

Overall, the results show that targeting isolation strategies based on individuals' spatial behaviour may reduce the number of Ebola infection cases, when isolating at least 15\% of the whole population. For smaller isolation percentages, no significant effect was observed.

More specifically, as probably expected, isolation based on $Progmosis$ risk model seems outperforming the other strategies. It shows significant effects on the reduction of the number of infections when isolating at least 15\% of the population. Its effectiveness is  due to the fact  this index combines individual information about user mobility with aggregated information about the outbreak itself. However, it is worth noting that the latter might not be easily available and, above all, reliable, especially in a developing country settings during an emergency. In these cases, computer-based simulations considering different estimations of the characteristics of the epidemics can prove useful, but with all the limitations associated to the modelling assumptions.

Similarly, the entropy strategy manages to delay the spreading, but at a lower extent when compared to the $Progmosis$ risk model. We observe similar effects only when we isolate a higher number of individuals (i.e., 30\%). Radius of gyration and home staying indexes lead to similar results. They are statistically significantly less effective than the entropy one, even though the gap in terms of performance is not substantial.

Given the well established link between the Shannon entropy of movements defined above and the heterogeneity of visitation and thus of contact patterns~\cite{song}, these results provide additional evidence of the significant impact of individuals' contact heterogeneity on the dynamics of an outbreak. Although, among mobility based targeting strategies, the entropy index seems more effective, all the three measures correlate with the heterogeneity of visitation patterns: the radius of gyration is a measure of the spatial dispersion of human movements. In general, we expect that individuals who have a large radius of gyration should be less predictable (i.e., high entropy). The home staying index, on the other hand, correlates with the spatial regularity of movements, so the lower the percentage of interactions the user had while he was at home the lower the regularity and the higher the heterogeneity of movements.

\section*{Discussion}
In this paper we have investigated the design and evaluation of targeted strategies for containing epidemic spreading considering the spatial properties of the population dynamics extracted from CDR data. We have explored and compared the effects of different measures for the identification of areas or individuals to be targeted.

We have focused on the case of person-to-person transmitted diseases, where social and environmental factors (e.g., crowded setting) are primary determinants of transmission. However, these factors are characterized by an intrinsic spatial variation, whose incorporation in epidemiological models remains a key theoretical challenge. Therefore, we have considered the problem of taking into consideration local spatial interactions and we have tried to capture and characterise the socio-geographical heterogeneity of transmission following two distinct approaches. 
Firstly, we have taken into consideration geographic heterogeneity, aiming at identifying \textit{geographic areas} with the higher opportunity of contact (i.e., where the majority of exchange is likely to originate). By exploiting the place rank measure for the definition of location accessibility and attractiveness, we have measured the ``spreading power'' of the nodes in a spatial network. By using this information, it is possible to rank and isolate nodes in order to contain the spreading of the epidemics.
Secondly, by considering spatial-based mobility indicators, we have quantified the ``spatial behaviour'' of \textit{single individuals} as a correlate of the contagion risk. Based on this, we have selected a subpopulation of individuals that is expected to become infected, and simultaneously infectious, with higher probability than the average population because of his/her mobility profile.

The results show the importance and effects of the spatial dimensions on the spreading of infectious diseases. While space influence has frequently reported anecdotally in the literature, there has been relatively little systematic investigation in this area. Our work tries to bridge this gap.
However, we are aware that this work has a series of limitations. The first is related to the assumption  concerning the reliability/validity of the epidemic model, which is fairly basic. However, we would like to underline the fact that the goal of this work was not in the definition of an accurate model of disease transmission, but on understanding the role of space in the design of countermeasures for containing epidemics spreading.

Another limitation is related to the data used for the experiment.  Although many studies have shown that mobile phone data provide a good proxy for human mobility studies~\cite{calabrese}, potential sources of inaccuracy do certainly exist. The first major concern, as only the mobile phone users are included in the data set, is a possible bias related to the specificity of the sample taken into consideration. The very large number of customers involved in this study (35\% of the whole population) seems to go against this specificity bias, even if there might be some bias related to the fact that we consider a single operator in this study. 
Other authors have proposed different models of human mobility patterns (see for example~\cite{matamalas}): Although the goal of this work is methodological, i.e., to propose a comparison of modeling different mitigation strategies considering the same underlying mobility model extracted from the CDRs, it would be interesting to investigate how different mobility models might affect our final results in terms of countermeasures? effectiveness. This is an issue that we plan to address in a future work.
There might also be a positive correlation between user mobility and communication frequency~\cite{iovan}: as billing records collect location only when a communication event occurs, a frequently moving (and calling) user has more location points than a more static one. So the movements of low-mobility users can be underestimated. However, it has been shown that in particular CDRs reproduce long-distance travel patterns with a high accuracy especially compared to transportation surveys~\cite{janzen}. For this reason, our research, founded on sub-prefecture flows, is probably less affected by this bias.

An additional and related concern is the sensitive nature of the data. The proposed approach (and, in particular, the individual-level isolation scenario) requires access to personal data. 
The access to this data without violating the personal right to privacy is a major concern~\cite{taylor}. Recent studies have tried to overcome the limits of a simple identifier re-coding or ``pseudo-anonymization''. For example, interesting approaches come from edge computing~\cite{garcialopez}. The idea is to pre-process the data directly on the device that produced it or by means of privacy-preserving machine learning techniques. More in general, the definition of a clear and ethical framework for this type of applications represents one of the major challenges for the application of models and technologies based on the analysis of mobile data.

\begin{backmatter}


\bibliographystyle{bmc-mathphys} 
\bibliography{ref}      



\section*{Figures}
%
\begin{figure}[h!] \captionsetup{width=.9\linewidth} 
    \includegraphics[width=.8\textwidth]{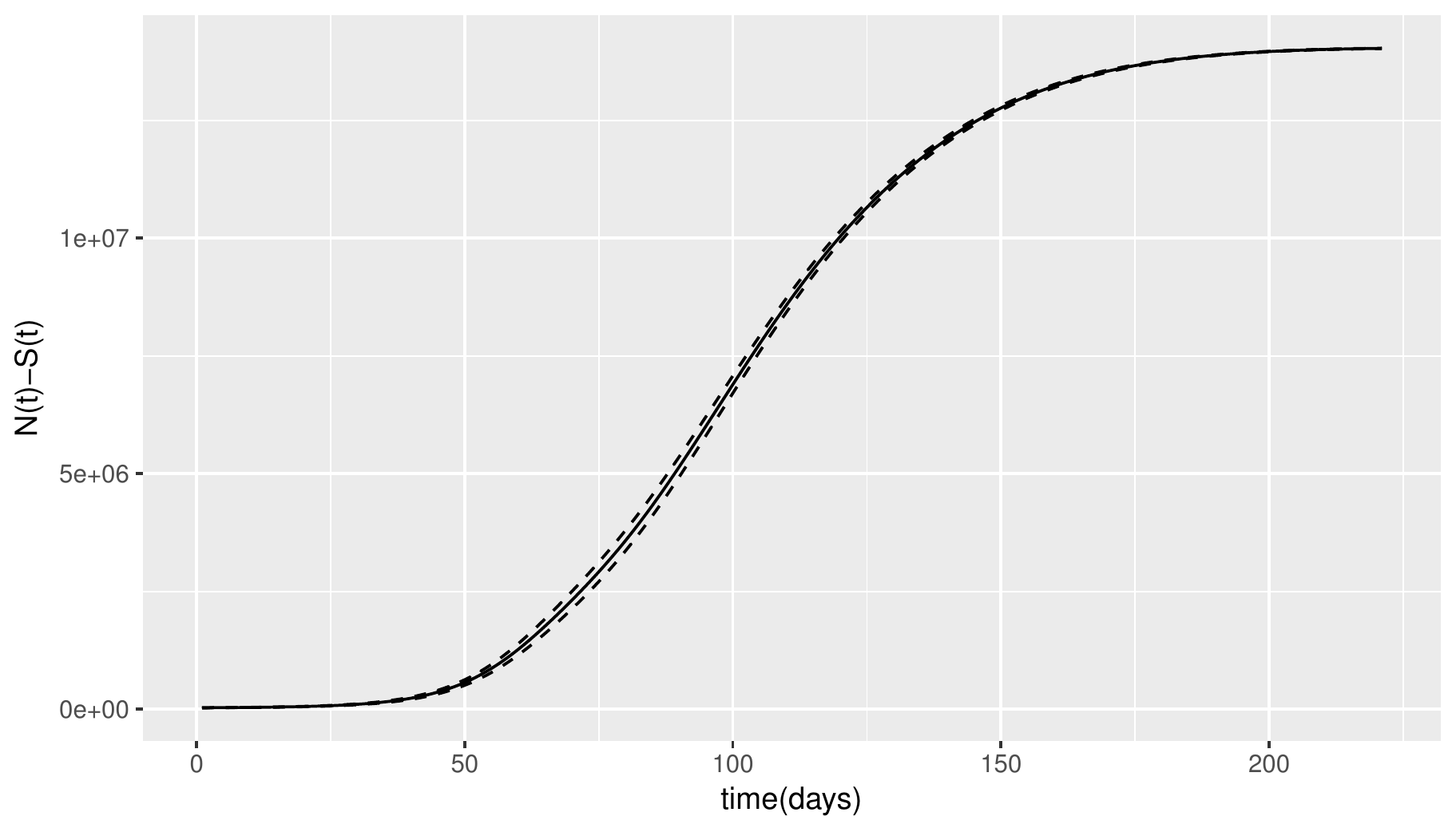} 
    \caption{\csentence{} Total number of infections since the beginning of the simulations, over a seven-month time period when no countermeasures are taken.}\label{figure2} 
\end{figure}
\begin{figure}[h!] \captionsetup{width=.9\linewidth} 
    \includegraphics[width=.8\textwidth]{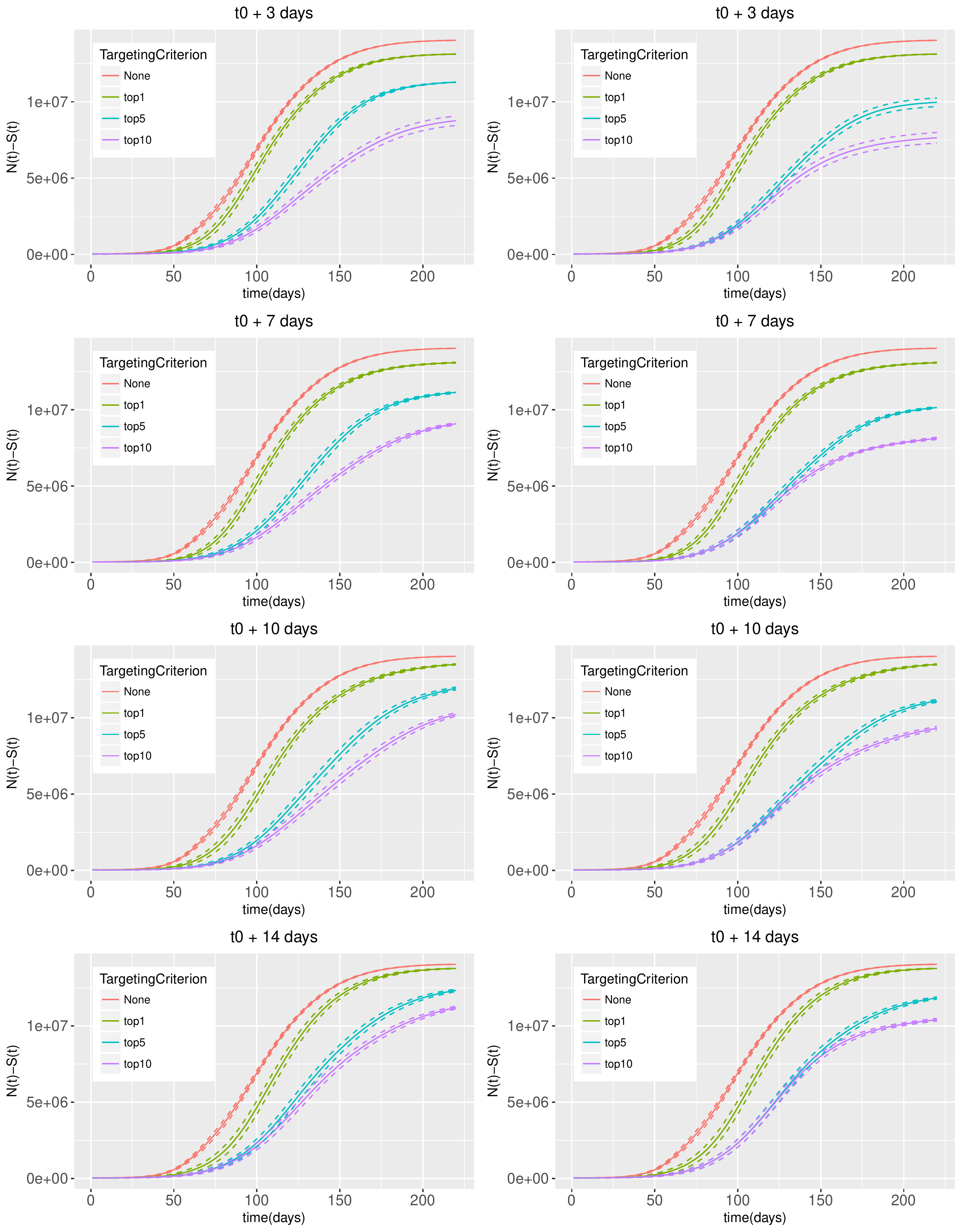} 
    \caption{\csentence{} Performance of sub-prefecture-level isolation based on two different strategies for determining targeted sub-prefectures, centrality-based (a), place rank-based (b). Solid lines represent the average number of infections over the time, dashed lines are the 95\% confidence interval. Interventions initiate three, seven, ten, fourteen days after the infection starts (i.e., t0)}\label{figure3} 
\end{figure}
\begin{figure}[h!] \captionsetup{width=.9\linewidth} 
    \includegraphics[width=.8\textwidth]{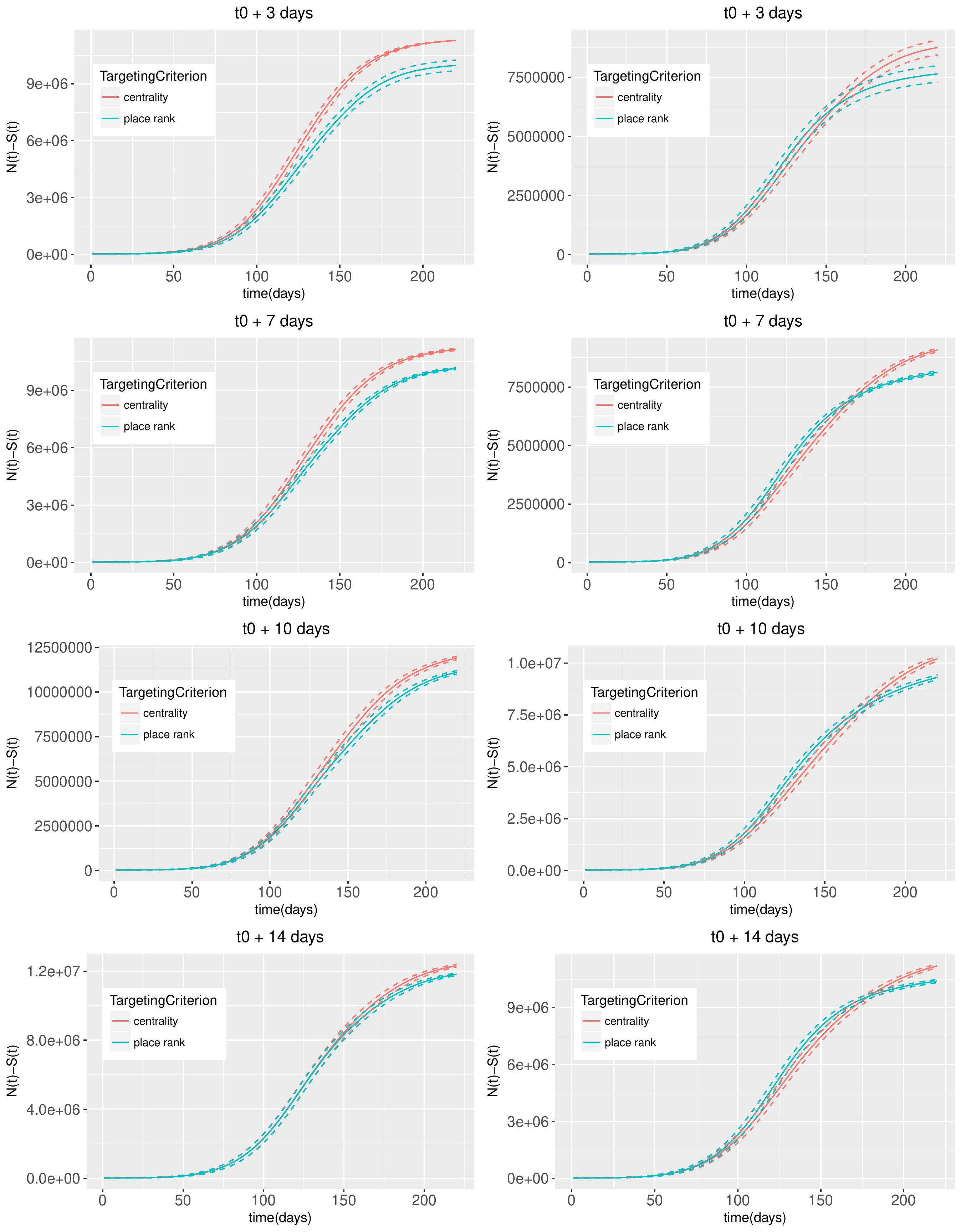} 
    \caption{\csentence{} Comparing place rank-based and centrality-based isolation when curbing the top 5 (a) and top 10 (b) highest risk sub-prefectures. Each panel shows the average number of infections over the time (solid lines), and the associated 95\% confidence interval (dashed lines). Interventions initiate three, seven, ten, fourteen days after the infection starts (i.e., t0)}\label{figure3.1} 
\end{figure}
\begin{figure}[h!] \captionsetup{width=.9\linewidth} 
    \includegraphics[width=.8\textwidth]{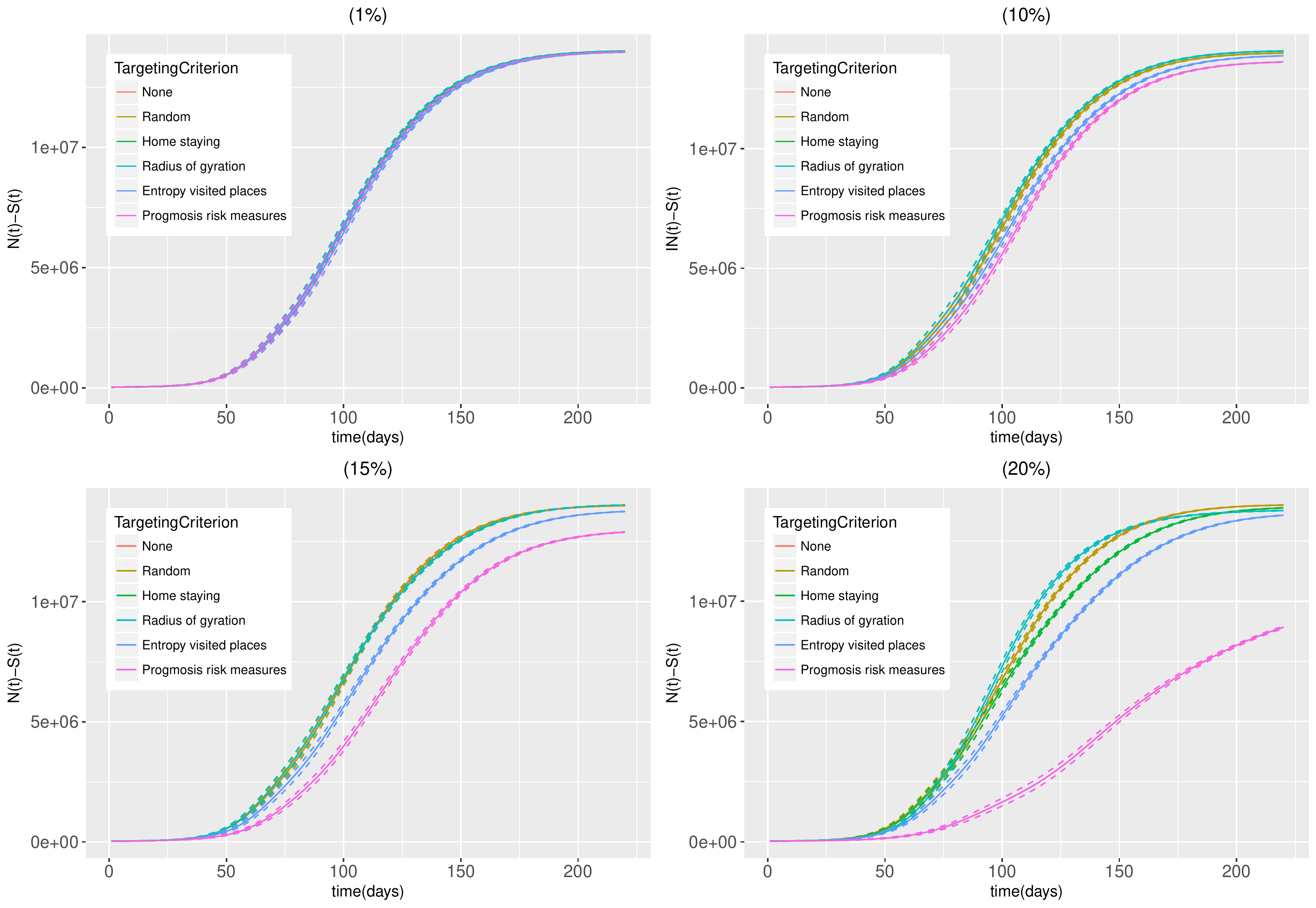} 
    \caption{\csentence{} Performance of individual-level isolation based on different spatial indexes for determining targeted individuals (\textit{none:} no counter measure, \textit{random:} a portion of individuals isolated randomly, \textit{radius of gyration:} a portion of individuals isolated based on the value of their radius of gyration, \textit{entropy visited places:} a portion of individuals isolated based on the value of the entropy of visited sub-prefectures, \textit{home staying:} a portion of individuals isolated based on the value of the percentage of time spent at home, \textit{Progmosis risk:} a portion of individuals isolated based on the value of the Progmosis risk model), The percentage of isolated individuals is set to 1\% (top-left), 10\%, 15\% and 20\% (bottom-right) of the whole population. Solid lines represent the average number of infections over the time, dashed lines represent the 95\% confidence interval.}\label{figure4} 
\end{figure}
%

\section*{Tables}
\begin{table}[h!]
\caption{Ebola specific parameters values.}
      \begin{tabular}{|cc|}
        $\beta$ &0.45 \\
        $\sigma$ & 0.18\\
        $\gamma$ &0.2\\
        $\rho$ &0.48\\ 
      \end{tabular}
       \label{table1}
\end{table}

\pagebreak
\setcounter{figure}{0}
\section*{Additional Files}
  \subsection*{Individual-based Targeting results}
\begin{figure}[h!] \captionsetup{width=.9\linewidth} 
    \includegraphics[width=.65\textwidth]{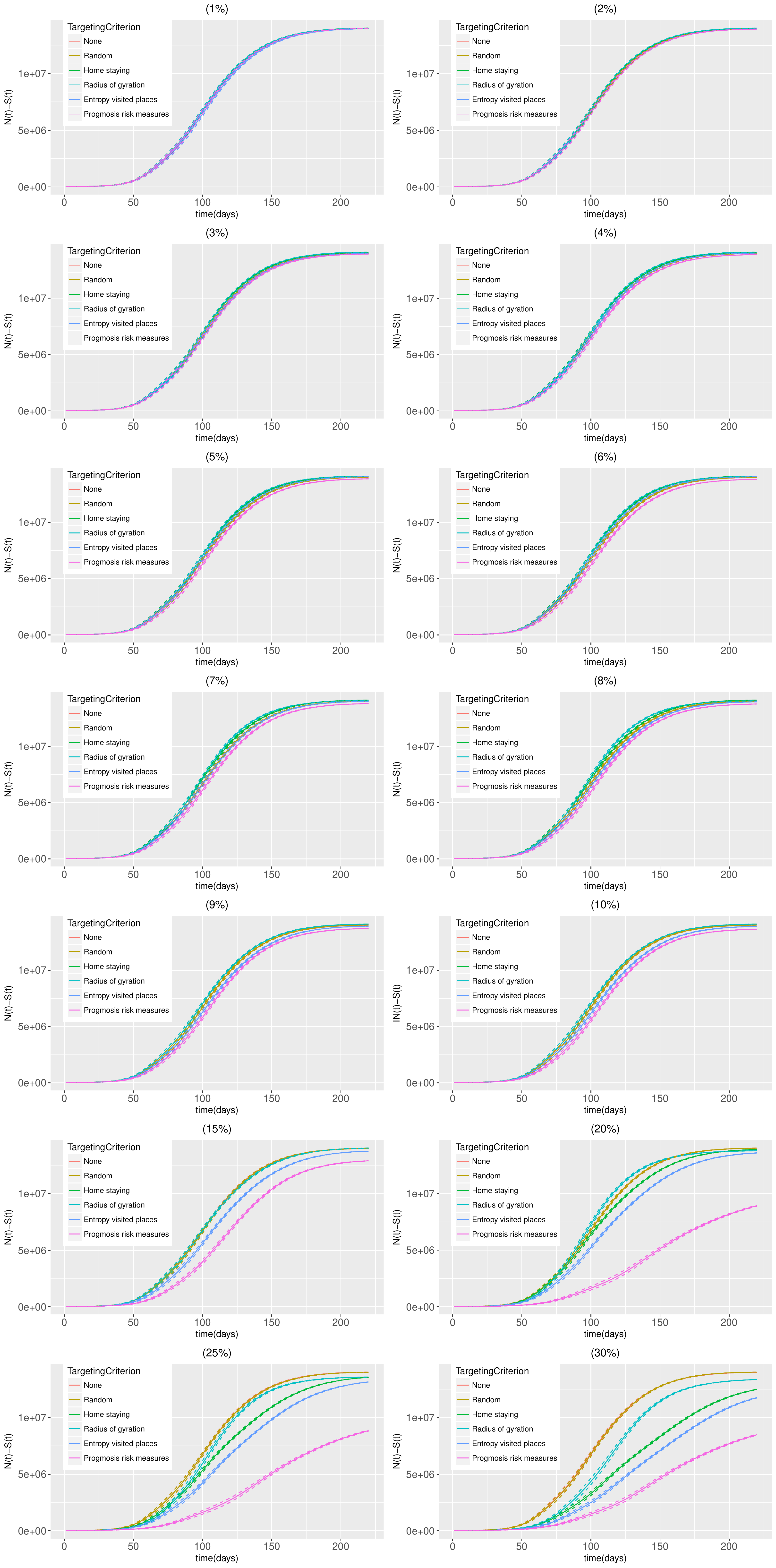} 
    \caption{\csentence{} Performance of individual-level isolation based on different spatial indexes for determining targeted individuals (\textit{none:} no counter measure, \textit{random:} a portion of individuals isolated randomly, 	\textit{radius of gyration:} a portion of individuals isolated based on the value of their radius of gyration, \textit{entropy visited places:} a portion of individuals isolated based on the value of the entropy of visited sub-prefectures, \textit{home staying:} a portion of individuals isolated based on the value of the percentage of time spent at home, \textit{Progmosis risk:}  a portion of individuals isolated based on the value of the Progmosis risk model), while varying the percentage of isolated individuals from 1\% (top-left) to 10\% of the whole population with step length 1, and from 10\% to 30\% (bottom-right) of the whole population with step length 5. Solid lines represent the average number of infections over the time, dashed lines represent the 95\% confidence interval.}\label{figure5} 
\end{figure}

\end{backmatter}
\end{document}